\documentclass{elsart}

\usepackage{natbib}
\usepackage{mathrsfs}
\usepackage{amssymb}
\usepackage{graphicx}
\usepackage{epstopdf}

\newcommand{\BibTeX}{ \textrm{B\kern-.05em\textsc{i\kern-.025em b}\kern-.08em
  T\kern-.1667em\lower.7ex\hbox{E}\kern-.125emX} }
\newcommand{\cm}{{\, \rm cm}}

\newcommand{\dyne}{{\, \rm dyne}}

\begin{document}

\begin{frontmatter}



\title{Elastic ice shells of synchronous moons:  Implications for cracks on Europa and non-synchronous rotation of Titan}

\author[caltech,ias]{Peter M. Goldreich},
\author[ucla,ias]{Jonathan L. Mitchell}

\address[caltech]{Caltech, Pasadena, CA 91125}
\address[ias]{Institute for Advanced Study, Princeton, NJ 08540}
\address[ucla]{Earth \& Space Sciences, Atmospheric \& Oceanic Sciences, and Institute for Geophysics \& Planetary Physics, University of California at Los Angeles, Los Angeles, CA 90095}

\begin{center}
\scriptsize
Copyright \copyright\ 2010 Peter M. Goldreich and Jonathan L. Mitchell
\end{center}
\begin{center}
\scriptsize
Icarus \emph{in press} (2010), doi:10.1016/j.icarus.2010.04.013
\end{center}



\begin{abstract}

A number of synchronous moons are thought to harbor water oceans
beneath their outer ice shells.  A subsurface ocean frictionally
decouples the shell from the interior. This has led to
proposals that a weak tidal or atmospheric torque might cause the
shell to rotate differentially with respect to the synchronously
rotating interior.  Applications along these lines have been made to
Europa and Titan.  However, the shell is coupled to the ocean by an
elastic torque. As a result of centrifugal and tidal forces,
the ocean would assume an ellipsoidal shape with its long axis aligned
toward the parent planet.  Any displacement of the shell away from
its equilibrium position would induce strains thereby increasing its
elastic energy and giving rise to an elastic restoring torque. 
In the investigation reported on here, the elastic torque is compared
with the tidal torque acting on Europa and the atmospheric torque
acting on Titan.  

Regarding Europa, it is shown that the tidal torque
is far too weak to produce stresses that could fracture the ice shell,
thus refuting an idea that has been widely advocated. Instead, it is
suggested that the cracks arise from time-dependent stresses due to
non-hydrostatic gravity anomalies from tidally driven, episodic
convection in the satellite's interior.

Two years of Cassini RADAR observations of Titan's surface have been
interpreted as implying an angular displacement of $\sim0.24$ degrees
relative to synchronous rotation. Compatibility of the amplitude and
phase of the observed non-synchronous rotation with estimates of the
atmospheric torque requires that Titan's shell be decoupled from its
interior. We find that the elastic torque balances the seasonal
atmospheric torque at an angular displacement $\lesssim0.05$ degrees,
effectively coupling the shell to the interior. Moreover, if Titan's surface
were spinning faster than synchronous, the tidal torque tending to
restore synchronous rotation would almost certainly be larger than
the atmospheric torque. There must either be a problem with the
interpretation of the radar observations, or with our basic understanding
of Titan's atmosphere and/or interior. 

\end{abstract}

\end{frontmatter}


\section{Introduction}

Synchronous, or near synchronous, spinning satellites with ice shells and subsurface oceans are an intriguing class of solar system bodies. Part of their appeal is the possibility that the oceans could support life. Currently Jupiter's satellite Europa and Saturn's satellite Titan are the prime candidates for membership in this class.

The induced magnetic field of Europa as measured by the magnetometer on the Galileo spacecraft essentially proves the presence of a current-carrying, near-surface liquid layer, i.e. a salty subsurface ocean \citep{2000Sci...289.1340K}.  All mechanisms proposed for creating the observed morphologies of cracks on Europa's surface require the presence of a near-surface fluid layer  \citep{1998Icar..135...25G,1999Icar..141..263G,1999Sci...285.1899H}. Most mechanisms also invoke an immeasurably slow, super-synchronous spin of the ice shell due to tidal torques associated with Europa's orbital eccentricity. The eccentricity, currently 0.01, is continually excited by resonant interactions with Io and Ganymede \citep{Peale_etal79, OjakangasStevenson86}. \cite{1984Icar...58..186G} suggest that tidal torques could
drive a super-synchronous rotation of Europa. \cite{1998Natur.391..368G} argue that the presence of a subsurface ocean might allow the surface ice shell to rotate non-synchronously even if the core is locked in synchronous rotation. 

Models of Titan's thermal evolution are consistent with the presence of a subsurface ocean. Moreover, motions of Titan's surface features as tracked by the Cassini spacecraft's radar have been interpreted as implying a slow, non- synchronous rotation \citep{Stiles_etal08,Stiles_etal09,Stiles_etal10}.   \cite{2008Sci...319.1649L} propose that the non-synchronous rotation is driven by a seasonally changing atmospheric torque acting on an ice shell decoupled from the interior by a subsurface ocean. 

The plan of our paper is as follows. As a result of centrifugal and tidal forces, the interiors of synchronous satellites have
ellipsoidal figures whose longest axes point toward the parent planet.  We derive these forces and resultant equilibrium shape in
\S  \ref{sec:potentialtheory}.  In \S  \ref{sec:energies}, we evaluate the increase in elastic and gravitational energy associated with the angular displacement of an elastic outer shell away from its equilibrium position and derive the elastic torque.  In \S \ref{sec:tidaltorque}, we outline a derivation of the tidal torque exerted on a synchronous shell due to the satellite's orbital eccentricity.  We derive the angular displacements at which the elastic torque balances the tidal torque acting on Europa's shell and the atmospheric torque acting on Titan's shell in \S \ref{sec:applications}.  The results of this exercise imply that the long cracks on Europa do not result from non-synchronous rotation of the ice shell driven by the tidal torque. They also cast doubt on claims of non-synchronous rotation for Titan.  We devote \S \ref{sec:gravityanomalies} to an explanation for the formation of large cracks on Europa's surface due to geoid variations associated with episodic mantle convection driven by tidal heating.  We end with a short discussion in \S\ref{sec:discussion}.  The main points of our investigation can be appreciated with order of magnitude estimates.  Consequently more detailed considerations are relegated to a series of Appendices.

\section{Potential theory}
\label{sec:potentialtheory}

This section contains background material selected for its relevance to our investigation. For details and derivations, the reader might consult \cite{2000ssd..book.....M}.

Consider a synchronously spinning satellite with an elastic outer shell, an underlying ocean, and a solid core.  Our focus is on the
elastic shell which has mean radius $R$ and thickness $d<< R$.  For clarity, the figures and Appendix \ref{app:alltorques} refer to a simpler model of an elastic shell and homogeneous fluid interior. 
Suppose the satellite moves on a circular orbit. Then the sum of the rotational and external tidal potential it feels is given by
\begin{equation}
V_{\rm{applied}} = \frac{n^2 r^2}{4} \left(\frac{10}{3}P_{2,0}(\cos\theta) - P_{2,2}(\cos\theta)\cos2\phi\right) \ ,
\label{eq:Vapplied}
\end{equation}
where $n$ is the orbital frequency, $P_{2,0}(x) = (3x^2-1)/2$ and $P_{2,2}(x) = 3(1-x^2)$.  To evaluate the satellite's shape under $V_{\rm{applied}}$, account must be taken of the potential, $V_{\rm{self}}$, due to the distorted satellite. Up to quadrupole order, for $r>R$,
\begin{equation}
V_{\rm{self}} = -\frac{GM}{r}\left\{1 + \left(\frac{R}{r}\right)^2
                      [C_{2,0}P_{2,0}(\cos\theta) + C_{2,2}P_{2,2}(\cos\theta)\cos2\phi]\right\} \ .
\label{eq:Vself}
\end{equation}
It is customary to define
\begin{equation}
J_2 \equiv -C_{2,0} \ .
\end{equation}
Expressed in terms of the principal moments of inertia, \begin{equation}
J_2=\frac{2C-A-B}{2MR^2} \quad\quad {\rm and} \quad\quad C_{2,2}=\frac{B-A}{4MR^2}\, .
\end{equation}

\begin{figure}[htbp]
\begin{center}
\includegraphics{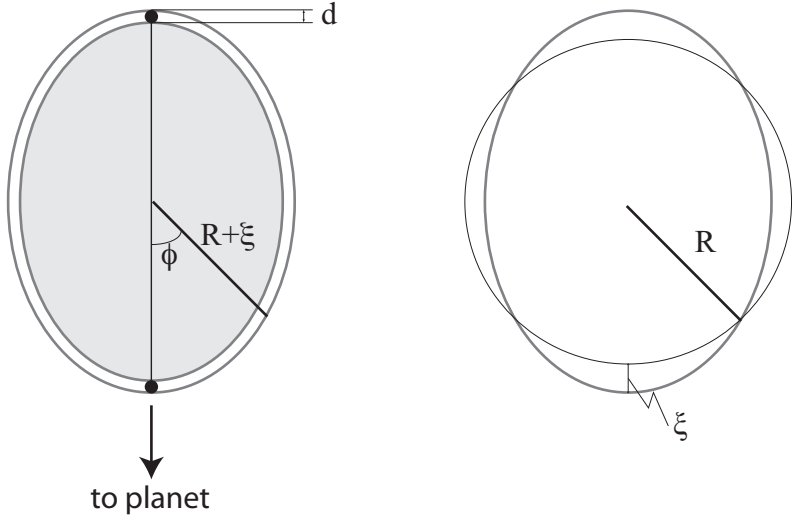}
\caption{Left:  Satellite with subsurface ocean. The outer shell (unfilled region) is elastic and the interior (gray region) is a fluid of uniform density.   The black dots are points of reference fixed to the shell.  Right:  Satellite shape is defined by the deviation, $\xi$, from the mean radius, $R$.}
\label{fig:icysatellite}
\end{center}
\end{figure}

Next we assume that the total potential, $V_{\rm{total}} = V_{\rm{self}} + V_{\rm{applied}}$ is constant on the satellite's surface. This
is appropriate if the deviatoric stress in the elastic shell vanishes. Expressing the shape of the surface by $R + \xi(\theta,\phi)$ (Figure \ref{fig:icysatellite}) and defining the parameter $q = n^2R^3/(GM)$, we arrive at
\begin{equation}
\frac{\xi(\theta,\phi)}{R} = \left(-\frac{5q}{6} + C_{2,0}\right)P_{2,0}(\cos\theta)
+ \left(\frac{q}{4}+C_{2,2}\right)P_{2,2}(\cos\theta)\cos2\phi \ .
\end{equation}
It is common practice to express the gravitational coefficients $C_{2,0}$ and $ C_{2,2}$ in terms of the Love number $k$ as
\begin{equation}
C_{2,0} = - \frac{5 k q}{6}, \ \ \ \ C_{2,2} = \frac{k q}{4} \ .
\end{equation}
The parameter $k$ depends upon the radial variation of density and elasticity. For bodies in
hydrostatic equilibrium, only the run of density matters. The fluid Love number is denoted by $k_f$ (see Appendix \ref{app:lovenumbers}). Assuming $k=k_f$, we obtain
\begin{equation}
\frac{\xi(\theta,\phi)}{R} = (1+k_f)\left(-\frac{5q}{6}P_{2,0}(\cos\theta)
                          + \frac{q}{4}P_{2,2}(\cos\theta)\cos2\phi \right) \ .
\label{eq:shape}
\end{equation}

\section{Energies of Orientation and the Elastic Torque}
\label{sec:energies}

\begin{figure}[htbp]
\begin{center}
\includegraphics{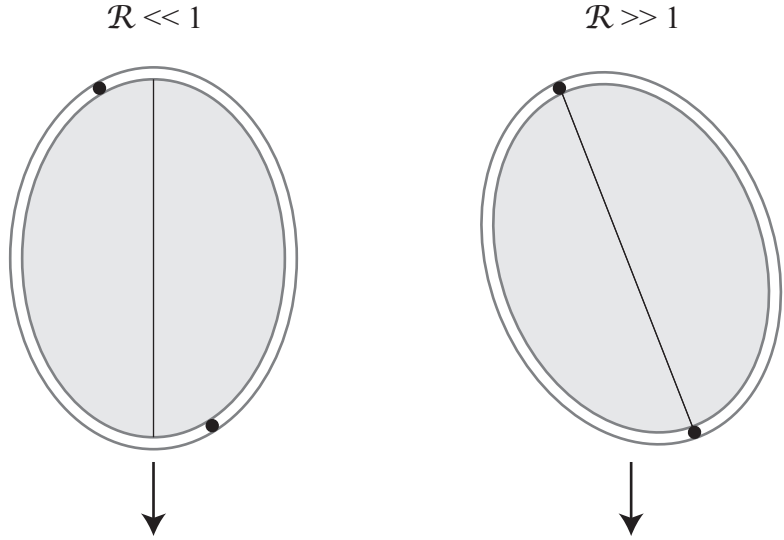}
\caption{Two limiting cases for a satellite rotated away from its equilibrium position.  The line indicates the long axis of the ellipsoidal figure.  If ${\mathcal R}<<1$, the shell deforms and stores elastic energy and the gravitational energy does not change.  If ${\mathcal R}>>1$, the shell behaves rigidly and the satellite's gravitational energy increases.  Europa and Titan are both in the former category, ${\mathcal R}<<1$ (see Table \ref{tab:parameters}).}
\label{fig:orientations}
\end{center}
\end{figure}

Consider a shell whose initial state is one of synchronous rotation and vanishing deviatoric stress.\footnote{The assumption of vanishing deviatoric stress might be challenged. This point is discussed in \S \ref{sec:discussion}.}
Suppose a spin-aligned torque applied to the shell rotates it by a small angle $\lambda<<1$.  Both elastic and gravitational energies increase. Two limiting cases are illustrated in Figure \ref{fig:orientations}.  In the first, the shell slides over the ocean which maintains its equilibrium shape. In the second, the entire satellite rotates with its shape remaining fixed.\footnote{In order for the shell not to touch the solid core, the ocean must be $\sim qR\sim \rm{few}\times10^2$ m or greater.  Titan's and Europa's oceans are estimated to be $\sim100-200$ km deep \citep{Sohl_etal03, Anderson_etal98}.}  In the simple example of a satellite with uniform density, only the elastic energy increases in the first case, whereas in the second only the gravitational energy does.

In the first case, the meridians of the shell are stretched and compressed due to the rotation giving rise to a strain $\sim \lambda q$.  The elastic energy of the shell after this rotation is $E_{\rm elas} \sim \mu \lambda^2q^2dR^2$ where $\mu$ is the shell's rigidity. 
In the second case, the equilibrium figure with a quadrupole $(B-A) \sim q MR^2$ is rotated against the gravitational torque of the planet, increasing the gravitational energy $E_{\rm grav} \sim q MR^2 n^2 \lambda^2\sim (GM^2/R)q^2 \lambda^2$.

The ratio of the energies in first and the second cases is ${\mathcal R} \equiv E_{\rm elas}/E_{\rm grav} \sim \mu d R^3/(G M^2)$.  Next we evaluate the numerical coefficients of these energies without assuming uniform density.

\begin{itemize}
\item
Expressions for the stress and strain tensors, $\sigma_{ij}$ and $u_{ij}$, are provided in Appendix \ref{app:stressstrain}.
The elastic energy density generated by rotating the shell an angle $\lambda$ with respect to the equilibrium figure of the ocean is \begin{equation}
{\mathcal E} = \sigma_{ij} u_{ij}/2\, .
\label{eq:elasNRGdensity}
\end{equation}
The total elastic energy
\begin{equation}
E_{\rm elas} = dR^2\int d\Omega {\mathcal E} =\frac{48\pi}{5}\left(\frac{1+\nu}{5+\nu}\right)(1+k_f)^2q^2\mu dR^2\lambda^2 \ ,
\label{eq:numericalelasNRG}
\end{equation}
with $\nu$ Poisson's ratio.
\item The increase in gravitational energy due to rotating the entire satellite with its shape fixed so that its long axis is displaced by an angle $\lambda$ from the direction toward the planet is
\begin{equation}
E_{\rm{grav}} = \frac{3}{2}k_f\frac{GM^2}{R}q^2\lambda^2 \ .
\end{equation}
\end{itemize}

The ratio \begin{equation}
{\mathcal R} = \frac{E_{\rm{elas}}}{E_{\rm{grav}}} = \frac{32\pi}{5}\left(\frac{1+\nu}{5+\nu}\right)\frac{(1+k_f)^2}{k_f}\frac{\mu dR^3}{GM^2}\ .
\label{eq:calR}
\end{equation}
In a real rotation the total increase of energy would be minimal. Thus if ${\mathcal R}<<1$, the shape of the shell would conform to that of the ocean's equilibrium figure,  whereas if ${\mathcal R}>>1$, the shell would maintain its shape and the ocean's figure would rotate with it.  For Europa and Titan, ${\mathcal R}<<1$ (Table \ref{tab:parameters}).  \citet{Karatekin_etal08, vanHoolst_etal08}; and \citet{vanHoolst_etal09} implicitly assume that the shell has infinite rigidity, or ${\mathcal R}>>1$.  Henceforth we take ${\mathcal R} <<1$.

The torque associated with the storage of elastic energy follows as
\begin{eqnarray}
\nonumber T_{\rm{elas}} &=& -\frac{\partial E_{\rm{elas}}}{\partial\phi} \\
                              &=& -\frac{96\pi}{5}\left(\frac{1+\nu}{5+\nu}\right)(1+k_f)^2q^2\mu dR^2\lambda \ . \label{eq:elastictorque}
\end{eqnarray}

A massless membrane provides the simplest model for a shell.  A rotated and deformed membrane exerts a pressure on the ocean's surface due to the hoop stress,
\begin{equation}
p_1=\left(\sigma_{\theta,\theta}+\sigma_{\phi,\phi}\right)\frac{d}{R}= -12 \left(\frac{1+\nu}{5+\nu}\right)(1+k_f)q\mu\frac{d}{R}\lambda\sin^2\theta\sin2\phi \, .
\end{equation}
The elastic torque arises from the back action of the ocean on the shell as indicated by Equation \ref{eq:pressuretorque}.  The primary effect on the ocean is to rotate the long axis of its ellipsoidal figure by an angle $\delta\sim {\mathcal R}\lambda$.\footnote{The change in ellipticity is of second order in $\delta$ and is ignored.}  In the simple case of a massless membrane, the planet's gravitational torque balances the pressure torque the shell exerts on the ocean. We evaluate $\delta$ for a more realistic model in Appendix \ref{app:surfacedistortion}.

The surface density perturbation that results from the deformation of the shell gives rise to an additional pressure torque $-(\rho g R/2\mu) T_{\rm{elas}}$ acting on the shell as shown in Equation \ref{eq:Tp2}.\footnote{The subscript $i$ denotes ice.}  This factor multiplying $T_{\rm{elas}} $ is of order unity for Europa and Titan. The minus sign follows because hoop stress is maximal where the shell is stretched thinnest.  It is noteworthy that the pressure torque on the shell due to the surface density perturbation is exactly canceled by the sum of the gravitational torques from the satellite and the planet on the distorted shell; we refer the motivated reader to Appendix \ref{app:pressuretorques} for a complete derivation.

\section{Tidal Torque on the Shell}
\label{sec:tidaltorque}

Tidal torques are often invoked to explain the origin of the putative non-synchronous rotation of Europa's ice shell.  A tidal torque acts on the synchronously rotating shell of a satellite that moves on an orbit of low eccentricity.  The torque arises from energy dissipated by the strain produced in the shell from diurnal tides.  A tidal wave can travel a distance $R$ across an ocean of depth $H$ in a time $R/\sqrt{gH}$.  This time is much shorter than the diurnal time of Europa and Titan for an ocean depth of 100 km.  The tidal response of the ocean is essentially instantaneous.  The free libration frequency of the shell (with restoring torque due to elasticity) is an order of magnitude less than the diurnal frequency, $n$.  Therefore to a good approximation, the shell remains stationary in the rotating frame and is deformed by the ocean tide below.  Time dependent components of the diurnal tide can be sorted into terms with inertial pattern speeds $\Omega_p = (n/2, 3n/2)$. 
Stresses and strains in the shell due to diurnal tides are given in Appendix \ref{app:stressstrain}. Corresponding elastic energies stored in the shell follow directly,
\begin{eqnarray}
E_{n/2} & = & \frac{3}{5}\pi\left(\frac{1+\nu}{5+\nu}\right)(h_t e q)^2\mu dR^2 \ , \\
E_{3n/2} & = & \frac{147}{5}\pi\left(\frac{1+\nu}{5+\nu}\right)(h_t e q)^2\mu dR^2 \ ,
\end{eqnarray}
where $h_t$ is the tidal Love number.  We expect $h_t<h_f$ since elastic stresses from diurnal tides do not have time to relax. A
mple expression for $h_t$ is given in Equation \ref{eq:tidalLove}. 
The torque exerted on the shell by each component tide is related to its rate of energy dissipation by $T = \dot{E}/(\Omega_p-n)$, where $\dot{E} = 2nE/Q$ follows from the definition of $Q$ as $2\pi$ times the peak stored energy divided by the energy dissipated over a cycle. Each component tide has frequency $n$ in the synchronous frame and the stored elastic energies are half the peak energy densities multiplied by the volume of the shell. Thus $T_{n/2} = - 4E_{n/2}/Q$, and $T_{3n/2} = 4E_{3n/2}/Q$, giving the total tidal torque
\begin{equation}
\label{eq:tidaltorque}
T_{e-tide} = \frac{576}{5}\pi\left(\frac{1+\nu}{5+\nu}\right)\frac{h_t^2}{Q}(qe)^2\mu dR^2 \ .
\end{equation}
The tidal torque tends to increase the spin rate because the tides are strongest near periapse where the orbital angular velocity exceeds the mean motion.

\section{Applications:  Europa and Titan}
\label{sec:applications}
We now apply the results derived in earlier sections to Europa and Titan.  We adopt parameter values $\mu = 4 \times 10^{10}$ dyne cm$^{-2}$, $\nu = 1/3$ and $Y=10^6$ dyne cm$^{-2}$ for ice.  The yield stress of ice, $Y$, is the most crucial and also the least well-constrained. We assume that the interiors of Europa and Titan have relaxed to hydrostatic equilibrium.  Models predict $k_f \approx 1$ for both bodies, a value that is consistent with gravity measurements made during Galileo flybys of Europa \citep{Anderson_etal98}.

We make use of four derived quantities.  The first two, $q$ and ${\mathcal R}$, were defined previously.  The third, the angle
$\lambda_{\rm{ext}}$ at which the elastic torque, Equation \ref{eq:elastictorque}, balances an externally applied torque, $T_{\rm{ext}}$, is
\begin{equation}
\lambda_{\rm{ext}} = \frac{5}{96\pi}\left(\frac{5+\nu}{1+\nu}\right)\frac{T_{\rm{ext}}}{(1+k_f)^2q^2\mu d R^2} \ .
\end{equation}
In the special case $T_{\rm{ext}} = T_{\rm{tidal}}$,
\begin{equation}
\lambda_t = \frac{6e^2}{Q}\left(\frac{h_t}{1+k_f}\right)^2 \  ,
\end{equation}
which is much smaller than the angle of geometrical libration.  In this case the thickness of the shell drops out.  The fourth quantity is the forced libration amplitude
\begin{equation}
\gamma_o = \frac{2\omega_o^2 e}{n^2-\omega_0^2} \ ,
\end{equation}
with eccentricity $e$ and (elastic) free libration frequency $\omega_o^2 = T_{\rm elas}/C_{\rm shell}$.\footnote{This amplitude and frequency is relevant for satellites with ${\mathcal R}<<1$.}  The moment of inertia of the shell $C_{\rm shell} = (8\pi/3)\rho R^4 d$ with $\rho$ the density of ice.

Parameters for Europa and Titan are given in Table \ref{tab:parameters}.

\begin{table}[htdp]
\caption{Parameters for Europa and Titan.}
\begin{center}
\begin{tabular}{l|ll}
		& 		Europa 			&      Titan \\ \hline \hline
$M$		& 	$4.8\times10^{25}$ g	&  	$1.4\times10^{26}$ g \\
$R$		& 	$1.6\times10^8$ cm		&  	$2.6\times10^8$ cm \\
$e$ 		& 		0.01				&      0.03   \\
$n$ 		&      $2\times10^{-5}$ s$^{-1}$ &      $4.6\times 10^{-6}$ s$^{-1}$   \\
$q$		& 	$5.4\times10^{-4}$		&      $3.9\times10^{-5}$ \\
${\mathcal R}$	& $2.1\times10^{-2}(d/10$km)	&      $0.1(d/100$km) \\
$\lambda_t$& $2.3\times10^{-4}/Q$	rad	&      $2\times10^{-3}/Q$ rad \\
$\lambda_a$& N/A					&      $7\times10^{-4}(d/100$km) rad\\
$\gamma_o$& $52''$				&      $5''$\\

\end{tabular}
\end{center}
\label{tab:parameters}
\end{table}%

\subsection{Europa}
${\mathcal R}<<1$ for Europa so to a good approximation its long axis points toward Jupiter.  The elastic torque balances the tidal torque at an angle $\lambda_t\sim 2.3\times 10^{-4}/Q$ radians.
We have assumed an $h_t$ defined in Equation \ref{eq:tidalLove}.  Rotation of the shell by this angle generates stresses of order
$q\mu\lambda_t\sim 5\times10^3/Q$ dyne cm$^{-2}$.\footnote{The maximum stress is a factor of 4.5 times larger.}
These stresses are clearly inadequate to crack the shell even for $Q=1$.  The thickness of the elastic shell does not enter into this calculation. 
Since for Europa $n>\omega_o$, the shell does not follow the tide and the shell's forced libration amplitude $\gamma_o \approx 52''$.

\subsection{Titan}
Titan possesses a thick atmosphere whose wind patterns exhibit
seasonal variations resulting in the exchange of angular momentum with
the surface.  Model estimates yield $\Delta L = 1.3\times10^{32}$ g
cm$^2$ s$^{-1}$ for the total seasonal angular momentum exchange
\citep{2005GeoRL..3224203T,2009ApJ...692..168M}.  The exchange occurs
at twice Saturn's orbital frequency, $\omega$, giving rise to a
maximum torque
\begin{equation}
T_{\rm{atmosphere}} \ = \ 2 \omega \Delta L \ \simeq \ 2\times10^{24} \ \rm{dyne \ cm}\ .
\end{equation}
This torque is capable of forcing a 100 km thick shell to rotate by
$\lambda_a=7\times10^{-4}$ radians relative to the equilibrium shape
of the ocean's surface. This angle is small in comparison to the
angular displacements of 0.12 deg/year reported by the Cassini radar
team \citep{Stiles_etal10}.  Moreover, since the shell is effectively
coupled to the interior, the angular displacement would be in phase
with the torque instead of half a cycle out of phase as it would be if
the shell could rotate independently of the interior.  This latter
point was raised by \cite{vanHoolst_etal09}, but they came to this
correct conclusion by incorrect arguments for reasons presented in \S
\ref{sec:energies}. 
Since for Titan $n>>\omega_o$, the shell does not follow the tide and the shell's forced libration amplitude $\gamma_0 \approx 5''$.

There is an additional problem with the notion that Titan's surface
is spinning faster than synchronous. If it were, tides raised by Saturn
on Titan would contribute an additional torque $\sim3\times10^{28}/Q$ dyne cm
tending to restore synchronous spin. Unless $Q>10^4$, a remarkably large value
for a solid body, the tidal torque would overwhelm the atmospheric torque.

\section{Mantle Convection, Gravity Anomalies, \& Cracks in Europa's Ice Shell}
\label{sec:gravityanomalies}

Having established  that  non-synchronous rotation did not cause the long cracks observed in  Europa's ice shell, our challenge is to come up with a more promising hypothesis.  Here we propose that like Io, Europa undergoes episodic tidal heating resulting in time-dependent, non-hydrostatic stresses which perturb its external gravitational potential.  The shape of the ocean surface and ice shell would respond to changes in the gravitational potential with cracks in the shell being a consequence of these changes of shape.

Two pieces of evidence point to episodic heating of Io.  Estimates of the current heat flow from its interior,  $P\approx 10^{14}$ watt, exceed the firm upper bound of $3.5\times 10^{13}$ watt on the time-averaged tidal heating rate of its interior (see \S \ref{app:bound}).  A unique feature of episodic tidal heating is that it can cause a temporary shrinking of Io's orbit, just what is found by \cite{2009Natur.459..957L} in their recent analysis of the measured orbital motions of Io, Europa, and Ganymede.

Europa presents a more problematic case for episodic heating than Io does. Observations do not offer evidence for either volcanism or an enhanced heat flow.  However, the paucity of impact craters on Europa's surface suggests that it was resurfaced as recently as $10^8$ years ago, perhaps as the result of an episode of enhanced tidal heating.  Europa has a larger orbit and a smaller mass and radius than Io. Each of these tends to reduce the rate at which tides dissipate energy in its interior relative to that of Io. Even so, based on a $Q=100$ and a tidal rigidity scaled to that of Earth's Moon, Europa's eccentricity damping timescale is only $3\times 10^8$ years.  Moreover, on average, Europa's orbital eccentricity is likely to be much larger than Io's as it is pumped by Europa's two-body mean motion resonances with the more massive satellites Io and Ganymede.  The findings from \cite{2009Natur.459..957L} suggest that at times Io must be deeper in resonance with Europa than it currently is. At such times Europa's orbital eccentricity would be larger than its present value.

\begin{figure}[htbp]
\begin{center}
\includegraphics{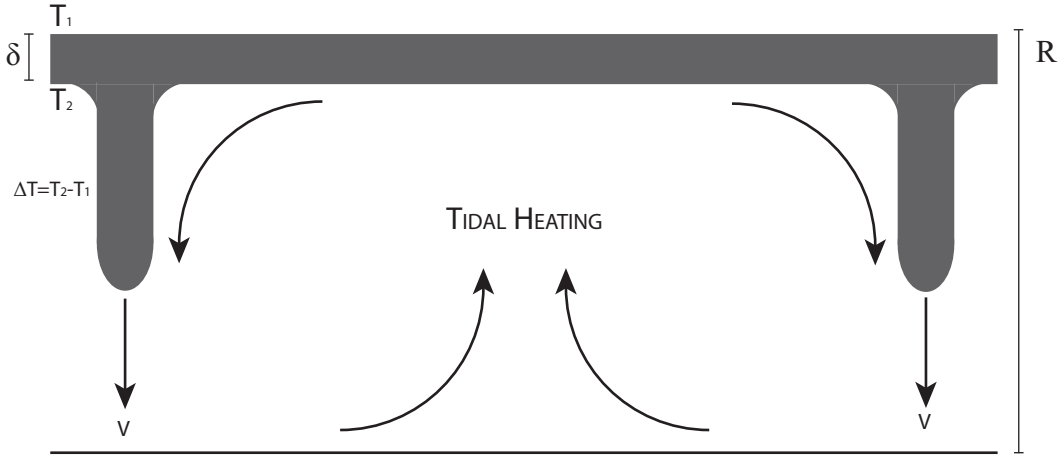}
\caption{Mantle convection.  A conducting lid of depth $\delta$ caps a convecting layer driven by negative buoyancy of the cooler lid.  The total depth of the mantle is $R$.  The temperature drops from $T_2$ to $T_1<T_2$ across the lid.  Density anomalies in the mantle from the subducting slabs vary on the convective time $R/v$.}
\label{fig:convection}
\end{center}
\end{figure}

Our proposal, although unavoidably speculative, is guided by the case for episodic tidal heating of Io made by \cite{OjakangasStevenson86}.  They investigate the response of Io to tidal heating based on detailed models for its rheology. Their calculations include, the angular momentum transfer from Jupiter that drives the expansion of Io's orbit, the excitation of Io's orbital eccentricity arising from its mean motion resonance with Europa, and the damping of the eccentricity due to  tides raised in the satellite by Jupiter.  In plausible scenarios, the satellite undergoes episodes of enhanced heating lasting of order $10^7$ years separated by quiescent intervals with durations of about $10^8$ years. Good matches are obtained for both the current values of Io's heat flow and orbital eccentricity.   
Now we investigate the consequences of assuming that Europa undergoes episodes of enhanced tidal heating similar to those proposed by \cite{OjakangasStevenson86} for Io.
Suppose just $10^{13}$ watt were dissipated and stored inside Europa during an interval of $10^7$ years.\footnote{About 10\% of the current heat flow from Io.}
If spread uniformly throughout the satellite, this heat would raise its internal temperature by $\Delta T\sim 50$ K and cause its radius to expand by $\Delta R/R\sim 10^{-3}$.\footnote{Europa's mantle accounts for $\sim$90\% of its radius \citep{Anderson_etal98}.} The tensile stress associated with the expansion would be of order $4\times 10^7\dyne\cm^{-2}$, well above what what is required to crack ice.  But tidal stresses are not spatially uniform so heterogeneous heating is a more realistic option.  It might generate convective plumes similar to the one illustrated in Figure 3. These would generate non-hydrostatic perturbations of the gravitational potential over a wide range of scales. 
The satellite's shell would adjust to an equilibrium shape in which the elastic and gravitational forces were in balance.  Next we provide an estimate of the scale dependent elastic stresses.

Consider an elastic shell whose relaxed state conforms to the hydrostatic shape of the synchronously rotating satellite.
The vertical displacement of the shell relative to the relaxed state is described by $\xi_\ell$, where $\ell$ denotes angular degree.
The displacement gives rise to strains that scale as
\begin{equation}
\epsilon_\ell\sim \left(1+\frac{\ell^2d}{R}\right)\frac{\xi_\ell}{R}\, .
\end{equation}
The first term in the parentheses arises from stretching and the second from bending. The latter is appropriate for a flat plate so its
dependence on $\ell$ might seem surprising. However $R/\ell$ is just the arc length associated with an angular separation
$1/\ell$ on a sphere of radius $R$.  The elastic energy is
\begin{equation}
E_{\ell, {\rm elas}}\sim \mu R^2d\epsilon_\ell^2\sim \mu d(1+\ell^2d/R)^2\xi^2_\ell\, .
\end{equation}
The gravitational energy associated with the departure of the ocean's surface from an equipotential is given by
\begin{equation}
E_{\ell, {\rm grav}}\sim g\rho R^2\left(\zeta_\ell-\xi_\ell\right)^2\, ,
\end{equation}
where $\zeta_\ell$ is the height of the geoid. The shell deforms to minimize the sum of the elastic and gravitational energies. Thus
\begin{equation}
\xi_\ell\sim \frac{\zeta_\ell}{1+{\mathcal R}(1+\ell^2d/R)^2}\, ,
\label{eq:xiell}
\end{equation}
and
\begin{equation}
\epsilon_\ell\sim \frac{(1+\ell^2 d/R)\zeta_\ell/R}{1+{\mathcal R}(1+\ell^2d/R)^2}\, .
\label{eq:epsell}
\end{equation}
Here ${\mathcal R}\ll 1$ is defined in Equation \ref{eq:calR}.

If the equipotential were to obey Kaula's rule \citep{1966tsga.book.....K}, which approximately characterizes the scale dependence of the surface gravitational potentials of Earth, Venus, Mars, and the Moon, the rms potential difference would vary in direct proportion to the inverse angular separation, $\Delta \phi \propto \ell^{-1}$. \footnote{At least for Earth and Venus, the low order components of the non-hydrostatic part of the gravitational potential are thought to arise from mantle convection.} 
The height of the geoid would scale similarly, $\zeta_\ell \sim \Delta\phi/g \sim \ell^{-1}$.  On the largest scales $\xi_\ell \sim \zeta_\ell$ and the shape of the shell would conform to that of the surface equipotential giving rise to strains that decrease as $\epsilon_\ell \sim \epsilon_o \ell^{-1}$ with increasing $\ell$.\footnote{Here $\epsilon_o$ is the strain at small $\ell$.}
For $\ell>(R/d)^{1/2}$, the elastic energy from bending the shell exceeds that from stretching.  In this regime the strain would increase linearly with $\ell$, $\epsilon_\ell \sim \epsilon_o (d/R)\ell$, up to a scale $\ell_M$ given by
\begin{equation}
\ell_M \sim {\mathcal R}^{-1/4} \left(\frac{R}{d}\right)^{1/2} \ .
\end{equation}
For $\ell>\ell_M$, the vertical displacement of the ice shell would be significantly smaller than the height of the geoid,
\begin{equation}
\frac{\xi_\ell}{\zeta_\ell} \sim \frac{1}{{\mathcal R}\ell^4}\frac{R^2}{d^2} \sim \left(\frac{\ell_M}{\ell}\right)^4 \, .
\end{equation}
Thus the strain would be given by
\begin{equation}
\epsilon_\ell \sim \frac{1}{{\mathcal R}^{1/2}}\left(\frac{\ell_M}{\ell}\right)^2\frac{\zeta_\ell}{R}\sim \frac{1}{{\mathcal R}}\frac{R}{d}\frac{\epsilon_o}{\ell^3}\, .
\end{equation}
The parameters listed in Table \ref{tab:parameters} for Europa yield $R^{3/2}/d^{1/2}\sim130$ km for the length scale at which the strain is minimal, and $R\ell_M\sim50$ km for the length scale at which the surface no longer conforms to the geoid.  Global cracks are the primary focus of our paper. However, we note that the scale of Europa's cycloidal cracks is similar to that at which strains due to stretching and bending have comparable magnitudes.  Moreover, an interplay between strains due to bending and stretching accounts for the formation
of cycloidal cracks when blunt objects tear thin, brittle sheets \citep{Ghatak_Mahadevan_03}.  Perhaps Europa's cycloids form from the blunt tearing action of $\sim$100 km scale, time-dependent, geoid anomalies.

\section{Discussion}
\label{sec:discussion}

We conclude with a few parting thoughts.

Our calculation of the elastic torque follows from the assumption that the shell's deviatoric stress vanishes in its tidally deformed state.   But there is another possibility.\footnote{As pointed out to us by both Re'em Sari and Jay Melosh.}   The shell might have formed rotating super-synchronously with a deviatoric stress that would vanish if it were to assume an axisymmetric shape. Then non-synchronous rotation could proceed without any variation of elastic energy and thus unimpeded by an elastic torque. Although this is a logical possibility, the weakness of the drivers of non-synchronous rotation for Europa and Titan render it implausible.  The tiny values of $\lambda_t$ calculated in \S \ref{sec:applications} are a measure of just how axisymmetric the unstressed shells would have to be in order to sustain non-synchronous rotation. 

We have demonstrated that Europa's long cracks are not opened by stresses due to non-synchronous rotation driven by the
torque arising from the diurnal tide.  Although this removes much of the {\it evidence} for non-synchronous rotation, it does not preclude it.  Provided the small stresses estimated in \S \ref{sec:applications} could relax on timescale $t_{\rm relax}$, nonsynchronous rotation {\it might} proceed on timescale $t_{\rm relax}/\lambda_t$. Our use of might instead of would is deliberate.
If the shell were to spin faster than synchronous, it would be subject to a component of tidal torque arising from the semi-diurnal tide which would tend to restore synchronous spin.  In the limit of small orbital eccentricity, the semi-diurnal component of tidal torque would overwhelm the diurnal component unless the tidal $Q$ were a steeply increasing function of period.

Non-hydrostatic moment differences would drive the ice shell to reorient with respect to a system of axes pointing toward the planet, along the orbit, and normal to it. However, because the ellipsoidal shape of the ocean surface would maintain its orientation in these axis, the elastic torque would oppose this type of polar wander \citep{Willemann84}. Significant reorientation of the shell would require that the ice either crack or creep.

Provided Titan's shell is $\sim100$km thick, the presence of an
elastic torque presents a challenge to the interpretation of Titan's
non-synchronous rotation.  Observational evidence for the
non-synchronous rotation of Titan cannot be lightly
dismissed. Although we suspect an error in the data analysis, it is
also possible that we are missing something important in our understanding
of the properties of Titan's atmosphere and/or interior.

\appendix*

\section{Appendix}
\subsection{Love numbers}
\label{app:lovenumbers}
Under the action of an applied potential $U$, the surface of a self-gravitating body of radius $R$ is lifted by $\xi = hU_s/g$, where $U_s$ is the external potential evaluated at the surface.  The distortion of the body gives rise to an additional gravitational potential $kU_s$.  Love numbers $h$ and $k$ are derived below for simple cases relevant to this paper.

The external potential acting on the satellite is taken to have angular dependence proportional to $P_{2,0}(\cos\gamma)$.  The response of the satellite can be written as a surface density perturbation of the form $\Sigma = \Sigma_0P_{2,0}(\cos\gamma)$.  Its potential is given by
\begin{equation}
U_{\Sigma} = \frac{4\pi GR\Sigma}{5} \ .
\end{equation}

\noindent \emph{Example 1:}  An incompressible fluid body of density $\rho$.  The surface is lifted by $\xi=\Sigma/\rho$.  Then from the definition of $h$,
\begin{equation}
\xi = \frac{U_s}{g}+\frac{3}{5}\xi \ \ \ \ \mbox{which implies} \ \ \ \ h = \frac{5}{2} \ \rm{\&} \ k=\frac{3}{2} \ .
\end{equation}

\noindent \emph{Example 2:}  A rigid sphere of density $\rho_c$ and radius $R_c$ overlain by an ocean of density $\rho $ and surface radius $R$.  Define $\overline{\rho} = [\rho_cR_c^3+\rho (R^3-R_c^3)]/R^3$.  Then
\begin{equation}
\xi = \frac{U_s}{g}+\frac{3\rho }{5\overline{\rho}}\xi \ \ \ \ \mbox{which implies} \ \ \ \ h = \frac{5\overline{\rho}}{5\overline{\rho}-3\rho } \ \rm{\&} \ k=\frac{3\rho }{5\overline{\rho}-3\rho } \ .
\label{eq:tidalLove}
\end{equation}

Under loading by a layer of surface density $\Sigma$, the boundary condition at the top of the ocean becomes $g\Sigma-\rho U={\rm const}$. Thus the distortion of the water's surface satisfies
\begin{equation}
\xi_w = -\frac{p}{g\rho } + \frac{4\pi GR\Sigma}{5g} + \frac{\rho }{\overline{\rho}}\xi_w = -\frac{\Sigma}{\rho } + \frac{3\Sigma}{5\overline{\rho}} + \frac{\rho }{\overline{\rho}}\xi_w \  ,
\end{equation}
which yields
\begin{equation}
\xi_w = - \frac{\Sigma}{\rho } \ .
\end{equation}

\subsection{Stresses and Strains}
\label{app:stressstrain}
We follow the technique described in \cite{MatsuyamaNimmo08} to calculate components of the stress tensor associated with
the reorientation of an elastic shell relative to the equilibrium shape of the ocean. \subsection{Stress From Rotation of Shell By Angle $\lambda$}

After factoring out the coefficient
\begin{equation}
\Gamma \equiv 3 \left(\frac{1+\nu}{5+\nu}\right)(1+k_f)q\mu\lambda \  ,
\label{eq:gamma}
\end{equation}
we find
\begin{eqnarray}
\nonumber \sigma_{\theta\theta} & = &  -\Gamma(2+\sin^2\theta)\sin2\phi \ , \\
\nonumber \sigma_{\phi\phi} & = &  \Gamma(2-3\sin^2\theta)\sin2\phi \ , \\
\nonumber \sigma_{\theta\phi} & = &  2\Gamma\cos\theta\cos2\phi \ , \\
\sigma_{\phi\theta} & = & \sigma_{\theta\phi} \, .
\label{eq:stresses}
\end{eqnarray}

\subsection{Time Variable Stress From Diurnal Tide}

After factoring out the coefficient
\begin{equation}{\mathcal A}\equiv \left(1+\nu\over 5+\nu\right)h_teq\mu\,
\end{equation}
we obtain: for pattern speed $n/2$,
\begin{eqnarray}
\sigma_{\theta\theta}&=&-\frac{3}{4}{\mathcal A}(2+\sin^2\theta)\cos(2\phi+nt)\, ,\cr \sigma_{\phi\phi}&=&\frac{3}{4}{\mathcal A}
(2-3\sin^2\theta)\cos(2\phi+nt)\, ,\cr  \sigma_{\theta\phi}&=&-\frac{3}{2}{\mathcal A}\cos\theta\sin(2\phi+nt)\, ,
\end{eqnarray}
and for pattern speed $3n/2$,
\begin{eqnarray}
\sigma_{\theta\theta}&=&\frac{21}{4}{\mathcal A}(2+\sin^2\theta)\cos(2\phi-nt)\, ,\cr \sigma_{\phi\phi}&=&-\frac{21}{4}
{\mathcal A}(2-3\sin^2\theta)\cos(2\phi-nt)\, ,\cr  \sigma_{\theta\phi}&=&-\frac{21}{2}{\mathcal A}\sin\theta\cos(2\phi-nt)\, .
\end{eqnarray}

\subsection{Strains from Stresses}

Components of the strain tensor follow from those of the stress according to
\begin{eqnarray}
\nonumber u_{\theta\theta} & = & \frac{\sigma_{\theta\theta} - \nu\sigma_{\phi\phi}}{2\mu(1+\nu)} \ , \\
\nonumber u_{\phi\phi} & = & \frac{\sigma_{\phi\phi} - \nu\sigma_{\theta\theta}}{2\mu(1+\nu)} \ , \\
\nonumber u_{\theta\phi} & = & \frac{\sigma_{\theta\phi}}{2\mu} \ , \\
u_{\phi\theta} & = & u_{\theta\phi} \ .
\label{eq:strains}
\end{eqnarray}

\subsection{Pressure Torques}
\label{app:pressuretorques}
The hoop stress from the deformed shell exerts a pressure perturbation, $p_1$,  on the ocean's surface;
\begin{equation}
\label{eq:hoopstress}
p_1 = (\sigma_{\theta\theta}+\sigma_{\phi\phi}) \frac{d}{R} = -12 \left(\frac{1+\nu}{5+\nu}\right)(1+k_f)q\mu\frac{d}{R}\lambda\sin^2\theta\sin2\phi \, ,
\label{eq:p1}
\end{equation}
with $\sigma_{\theta\theta}, \ \sigma_{\phi\phi}$ given in Equation \ref{eq:stresses} and values of $k_f$ derived in Appendix \ref{app:lovenumbers}.
Acting on the shell, $p_1$ produces a torque $T_{p_1}$;
\begin{equation}
\label{eq:pressuretorque}
T_{p_1} = -R^2 \int d\Omega\frac{\partial\xi}{\partial\phi} p_1\, ,
\label{eq:Tp1}
\end{equation}
where
\begin{equation}
\frac{1}{R}\frac{\partial\xi}{\partial\phi} = - \frac{3}{2} (1+k_f)q\sin^2\theta\sin2\phi\, \end{equation}
and $\xi$ is given in Equation \ref{eq:shape}.
It is a simple exercise to demonstrate that $T_{p_1}=  T_{\rm{elas}}$ (Equation \ref{eq:elastictorque}).

We neglect the difference in the densities of water and ice, $\rho_w=\rho_i=\rho$.  Perturbations in the surface density of the shell, $\Sigma_2=-d\rho(u_{\theta\theta}+u_{\phi\phi})$, from strains in Equation \ref{eq:strains} are responsible for an additional pressure perturbation, $p_2$;
\begin{equation}
p_2 =-g \Sigma_2\,  .
\label{eq:p2}
\end{equation}
Note that
\begin{equation}
\frac{p_2}{p_1} = - \left(\frac{1-\nu}{1+\nu}\right)\frac{g\rho R}{2\mu} \ ,
\end{equation}
so
\begin{equation}
T_{p_2} = - \left(\frac{1-\nu}{1+\nu}\right)\frac{g\rho R}{2\mu} T_{\rm{elas}} \ .
\label{eq:Tp2}
\end{equation}
Since $g\rho R/2\mu \sim 1$ for Europa and Titan, the shell is more than a massless membrane and $T_{p_2}$ cannot be neglected.

The gravitational torque on the surface density perturbation $\Sigma_2$ is
\begin{equation}
T_{g_2} = -R^2\int d\Omega\frac{\partial V_{\rm{total}}}{\partial\phi} \Sigma_2 \ ,
\end{equation}
where
\begin{equation}
\frac{\partial V_{\rm{total}}}{\partial\phi} = \frac{3GM}{2R}(1+k_f)q\sin^2\theta\sin2\phi = -g\frac{\partial\xi}{\partial\phi} \ ,
\end{equation}
and $V_{\rm total}$ is the sum of Equation \ref{eq:Vapplied} and Equation \ref{eq:Vself}.  $T_{g_2}$ exactly cancels $T_{p_2}$.\footnote{A fraction $1/(1+k_f)$ of $T_{g_2}$ is due to the planet and the remainder is due to the tidally distorted satellite.} 
\subsection{Distortion of the Surface}
\label{app:surfacedistortion}
The pressure perturbation, $p_1+p_2$ given in Equations \ref{eq:p1} and \ref{eq:p2}, distorts the shape of the ocean's surface by $\alpha$;
\begin{equation}
\label{eq:alpha}
\frac{\alpha}{R} = -\frac{h_t p_1+p_2}{\rho g R} =12 \left[h_t - \left(\frac{1-\nu}{1+\nu}\right)\frac{\rho gR}{2\mu}\right]\left(\frac{1+\nu}{5+\nu}\right)\frac{(1+k_f)q\mu d}{\rho  gR^2}\lambda\sin^2\theta\sin2\phi \ .
\end{equation}
Note that $\alpha$ is an order $d/R$ smaller than $\xi$, Equation \ref{eq:shape}.  The long axis of the ellipsoidal figure rotates by an angle
\begin{equation}
\frac{\delta}{\lambda} = 8 \left[h_t-\left(\frac{1-\nu}{1+\nu}\right)\frac{\rho gR}{2\mu}\right]\left(\frac{1+\nu}{5+\nu}\right)\frac{\mu d}{\rho  g R^2} \ .
\label{eq:delta}
\end{equation}
Two points are worth noting.
\begin{itemize}
\item $\delta/\lambda$ is of order ${\mathcal R}$ which is small in applications we are considering.
\item The fractional change in ellipticity is of order $\delta$.
\end{itemize}

\subsection{Tying It All Together}
\label{app:alltorques}

We offer an overview of the torques acting on different parts of the satellite and show that they sum to zero. To maintain clarity, this is done in the context of a simple model.  The satellite is taken to move on a circular orbits. It consists of an elastic ice shell and a fluid water interior. The deviatoric stresses vanish in its initial state. The density difference between unstressed ice and liquid water is neglected so $k_f = 3/2$ and $h_t = 5/2$. 
Imagine that a torque $T_{\rm{applied}} = -T_{\rm{elas}}$ (Equation \ref{eq:elastictorque}) is applied to the elastic shell which rotates by a small angle, $\lambda$, with respect to the equilibrium shape of the ocean that lies underneath it. The ocean's long axis rotates relative to the direction to the planet by the smaller angle $\delta$ (Equation \ref{eq:delta});
\begin{equation}
\frac{\delta}{\lambda} = 20 \left[1 - \frac{2}{5}\left(\frac{1-\nu}{1+\nu}\right) \frac{\rho g R}{2\mu}\right] \left(\frac{1+\nu}{5+\nu}\right) \frac{\mu d}{\rho g R^2} \ .
\end{equation}
Stemming from its rotation by $\delta$, the planet exerts a torque on the ocean
\begin{equation}
T_{\delta} = - 3(B-A)n^2\delta = - \frac{9GM^2}{2R}q^2\delta = \left[1-\frac{2}{5}\left(\frac{1-\nu}{1+\nu}\right)\frac{\rho gR}{2\mu} \right]T_{\rm{elas}} \ .
\end{equation}
It also exerts a torque on the perturbed surface density of the shell given by
\begin{equation}
T_{g_{21}} = \frac{2}{5}\left(\frac{1-\nu}{1+\nu}\right) \frac{\rho gR}{2\mu} T_{\rm{elas}} \ .
\label{eq:Tg21}
\end{equation}
Summing these two torques confirms that the total planet torque equals $T_{\rm{elas}}$.

The torques on the shell also must sum to $T_{\rm{elas}}$.  From the back reaction of the pressure it exerts on the ocean below, there is
\begin{equation}
T_{p_1} = T_{\rm{elas}}
\end{equation}
from the hoop stress (Equation \ref{eq:Tp1}), and
\begin{equation}
T_{p_2} = - \left(\frac{1-\nu}{1+\nu}\right)\frac{g \rho R}{2\mu}T_{\rm{elas}}
\end{equation}
from the perturbed weight of the shell (Equation \ref{eq:Tp2}).  The gravitational torque includes
\begin{equation}
T_{g_{21}} = \frac{2}{5}\left(\frac{1-\nu}{1+\nu}\right) \frac{g\rho R}{2\mu}T_{\rm{elas}}
\end{equation}
from the planet acting on the shell's perturbed surface density (Equation \ref{eq:Tg21}), and
\begin{equation}
T_{g_{22}} = \frac{3}{5}\left(\frac{1-\nu}{1+\nu}\right)\frac{g\rho R}{2\mu} T_{\rm{elas}}
\end{equation}
from the fluid interior acting on the shell's perturbed surface density.  Summing the torques on the shell we have accounted for so far, we obtain
\begin{equation}
T_{\rm{shell}} = T_{p_1}+T_{p_2}+T_{g_{21}}+T_{g_{22}} = T_{p_1} = T_{\rm{elas}} \ .
\end{equation}
\label{lastpage}

\subsection{Bounding Tidal Dissipation Within Io, Europa \& Ganymede}
\label{app:bound}

The angular momentum $L$ and energy $E$ of a satellite with mass $M$ moving on a circular orbit with mean motion $n$ around a planet of mass $M_J$ may be written as
\begin{equation}
L=\frac{\mu^{2/3}M}{n^{1/3}} \quad {\rm and} \quad E=-\frac{\mu^{2/3}Mn^{2/3}}{2}\, ,
\end{equation}
where $\mu\equiv GM_J$.  We express the tidal torque the planet exerts on a satellite by
\begin{equation}
T=KM^2n^4\, ,
\end{equation}
with $K$ is assumed to be independent of both $M$ and $n$.

Now consider three satellites, denoted 1, 2, 3 in order of increasing
orbital semi-major axis. Conservation of angular momentum implies
\begin{equation}
{\dot L}_1+{\dot L}_2+{\dot L}_3=T_1+T_2+T_3.
\label{eq:LTbasic}
\end{equation}
If tides raised within the satellites maintain their orbital eccentricities at low levels, the total (time-averaged) power dissipated within them is given by the difference between the rate at which work is done by  tidal torques and that at which the orbital energy increases;
\begin{equation}
P=n_1T_1+n_2T_2+n_3T_3-{\dot E}_1-{\dot E}_2-{\dot E}_3.
\label{eq:Pbasic}
\end{equation}

Next we assume that the orbits are locked together by mean motion resonances such that
$n_1=2n_2=4n_3$ and that we can neglect $T_2$ and $T_3$ with respect to $T_1$. \footnote{Both assumptions
are valid for the Galilean satellites.}
Combining equations (\ref{eq:Pbasic}) and (\ref{eq:LTbasic}) then yields
\begin{equation}
P=KM_1^2n_1^5\left(1-\frac{\left[1+2^{-2/3}(M_2/M_1)+2^{-4/3}(M_3/M_1)\right]}{\left[1+2^{1/3}(M_2/M_1)+2^{2/3}(M_3/M_1)\right]}
\right)
\end{equation}
\cite{Lissauer_etal84} derive a similar result for the heating of Enceladus from interactions with Janus and Saturn's rings.
An upper bound on $K$ is obtained by assuming that the resonance formed early and that
subsequently the system expanded by a significant fraction of its current size during the age of the solar system, $t_{ss}$.  With these assumptions we find
\begin{equation}
K_{\rm ub}=\frac{GM_J}{13M_1a_1t_{ss}n_1^5}\left[1+2^{1/3}(M_2/M_1)+2^{2/3}(M_3/M_1)\right]
\end{equation}
Consequently,
\begin{equation}
P_{\rm ub}=\frac{2^{1/3}GM_J}{26a_1t_{ss}}\left(M_2+\frac{3}{2^{2/3}}M_3\right).
\end{equation}
With values appropriate to the Galilean satellites, $P_{\rm ub}=3.3\times 10^{13}$ watt.

\ack
We thank Jay Melosh and an anonymous referee for helpful comments, Zane Selvans for informative conversations and David Stevenson for wise counsel. JLM was supported by the W.M. Keck Foundation at the Institute for Advanced Study.

\end{document}